# PRE-PRINT

## Multi-Length-Scale Dopants Analysis of an Image Sensor via Focused Ion Beam-Secondary Ion Mass Spectrometry and Atom Probe Tomography

Bavley Guerguis[1,*], Ramya Cuduvally[1,2], Alexander Ost[3], Morvarid Ghorbani[1], Sabaa Rashid[2], Wilson Machado[4], Dan McGrath[4], Chris Pawlowicz[4], Brian Langelier[1,2], Nabil Bassim[1,2]

[1]*Department of Materials Science and Engineering, McMaster University, Hamilton, Ontario L8S 4L7, Canada*
[2]*Canadian Centre for Electron Microscopy, McMaster University, Hamilton, Ontario L8S 4M1, Canada*
[3]*Raith GmbH, 44263 Dortmund, Germany*
[4]*TechInsights, Ottawa, Ontario K2K 2X2, Canada*

***Corresponding author:** guerguib@mcmaster.ca

Keywords: Atom Probe Tomography, Focused Ion Beam Secondary Ion Mass Spectrometry, Semiconductors, Dopants

# PRE-PRINT

## Abstract

The following article presents a multi-length-scale characterization approach for investigating doping chemistry and spatial distributions within semiconductors, as demonstrated using a state-of-the-art CMOS image sensor. With an intricate structural layout and varying doping types/concentration levels, this device is representative of the current challenges faced in measuring dopants within confined volumes using conventional techniques. Focused ion beam-secondary ion mass spectrometry is applied to produce large-area compositional maps with a sub-20 nm resolution, while atom probe tomography is used to extract atomic-scale quantitative dopant profiles. Leveraging the complementary capabilities of the two methods, this workflow is shown to be an effective approach for resolving nano- and micro- scale dopant information, crucial for optimizing the performance and reliability of advanced semiconductor devices.

## Introduction

The characterization of dopants and their spatial distributions within confined volumes is a particularly challenging task in semiconductor metrology. Not only is such knowledge critical for optimizing the performance/reliability of micro-devices (Anon, 2023), but also for building a thorough understanding of material-electrical/optical properties (Takamizawa et al., 2012) and in failure analyses (Cuduvally et al., 2023).

Atom probe tomography (APT) has rapidly emerged as a promising technique for this purpose, offering 3D chemical mapping with an atomic-scale spatial resolution and a detection limit on the order of several atomic parts per million (Gault et al., 2021). However, routine application of APT for semiconductors has, to some degree, been hindered by compositional biases (Guerguis et al., 2024), field evaporation (FE) artifacts (Melkonyan et al., 2017), and the notably complex sample preparation (Thompson et al., 2007) (i.e., the region-of-interest (ROI) must be precisely located at the apex of a needle-shaped specimen with a diameter that is less than 100 nm). While no other technique matches APT's combined sensitivity and spatial resolution, these limitations emphasize correlative approaches as the optimal strategy for dopants analysis (e.g., supplementing/complementing APT data, guiding sample preparation, etc.).

A technique commonly employed alongside APT is (scanning)-transmission electron microscopy ((S)TEM), which is used to access structural, crystallographic, and phase information (Da Costa et al., 2024; Stoffers et al., 2017; Langelier et al., 2017), probe the FE mechanism (Dialameh et al., 2024; Fleischmann et al., 2018), and calibrate reconstructions (Mouton et al., 2019; Fougerouse et al., 2022; Ceguerra et al., 2021). Compositional information can be obtained from (S)TEM-based spectroscopy methods (i.e., energy dispersive X-ray spectroscopy (EDS/EDX) or electron energy loss spectroscopy (EELS)), but these do not possess the required sensitivity to resolve the minute dopant concentrations in state-of-the-art devices (Gault et al., 2021). Atomic force microscopy (AFM)-based techniques are also routinely applied; for instance, scanning capacitance microscopy (SCM) and scanning microwave impedance microscopy of capacitive response (SMIM-C) (Li & Ferrari, 2018). Still, these provide no direct chemical information, and require the use of carefully designed calibration standards, along with appropriate physical models, to yield quantitative measurements (Marchiando et al., 2000). Depending on the material system and dopant profile, this may not be possible, or particularly difficult.

In semiconductor metrology, secondary ion mass spectrometry (SIMS) is a well-established and one of the most widely-used techniques for chemical surface analysis, on accounts of its parts per billion (or higher) sensitivity (Stevie, 2015). With a primary ion beam size typically ranging from a few hundred nanometers to several micrometers, conventional SIMS averages composition over a relatively large area (e.g., 100×100 $\mu m^2$) and thus, its utility is severely limited for the analysis of individual devices/nanostructures. 'Self-focusing' SIMS has been proposed as a possible technique to overcome this limitation, exploiting the mechanism by which cluster ions form during analysis to access localized compositional detail (Franquet et al., 2016). However, this approach has a critical reliance on robust calibration curves for accurate quantification and is heavily influenced by slight inhomogeneities or defects. Two variants of conventional SIMS that aim to resolve this limitation are Nanoscale-SIMS (Li et al., 2020) (NanoSIMS) and focused ion beam-SIMS (Levi-Setti & Fox, 1980; Pillatsch et al., 2019) (FIB-SIMS), which leverage the use of a finer ion probe to substantially enhance the lateral resolution and enable site-specific measurements (i.e., nanoscale imaging or depth profiling). Depending on the acquisition parameters and the material system, a resolution of less than 50 nm can be achieved (De Castro et al., 2022) with a moderate detection limit (i.e., parts per million range (Whitby et al., 2012)).

Herein, a multi-length-scale characterization workflow is outlined, exploiting the complementary capabilities of FIB-SIMS and APT to conduct a detailed dopants analysis of a microelectronic device. The efficacy of such a methodology has been previously illustrated as in investigations of alloys (Hu et al., 2019; Pedrazzini et al., 2018; Seol et al., 2011, 2013; Da Rosa et al., 2020), nuclear (Wang et al., 2016; Kautz et al., 2021) and geological (Rickard et al., 2016) materials. FIB-SIMS addresses the need for large-area, high-resolution dopant mapping, offering a framework for investigating elemental distributions and profiles at the microscale (Stevie et al., 1999). APT may then be used to elucidate quantitative, nanoscale information from a subset of the device, as necessary.

## Materials and Methods

The sample used in this work is a back-illuminated CMOS image sensor with a 0.6 µm pixel array (Figure 1(A)), representative of modern optoelectronic architectures. High-angle annular dark field (HAADF)-STEM cross-section imaging (Figure 1(B); acquired on a Thermo Scientific Talos 200X with a camera length of 205 mm and a convergence angle of 22.5 mrad) shows the structural layout of the device, while the electronic structure, in terms of the relative doping types and concentration levels, can be resolved in correlating this with SCM and SMIM-C maps (Figure 1(C)) (Li & Ferrari, 2018). The contrast observed in the HAADF image reflects variations in elemental composition, with brighter regions corresponding with higher atomic number elements. The pixels, which have a depth of approximately 4 μm, are arranged periodically and separated by *p*-type front-deep trench isolation (F-DTI) filled with *p*-type doped polysilicon structures, interposed between metallic contacts at the top and microlenses below. A graded doping profile within the pixels, transitioning from *n*-type to *p*-type, is evident by the SCM map.

TEM-EDS was also performed (Figure 1(D)), with major elements characterized. Three dopants (i.e., B, P, and As) were identified based on prior conventional SIMS depth profiling (not shown), but as anticipated, these were not detected with TEM-EDS. For instance, the spatial distribution of B in Figure 1(D) shows no discernible pattern and closely resembles the background. Nevertheless, TEM-EDS does provide insight into the chemical structure, with O forming the F-DTI, W and Ti constituting the metallic contacts, and an organic composition for the microlenses. STEM-EELS was also applied (not shown), revealing similar results.

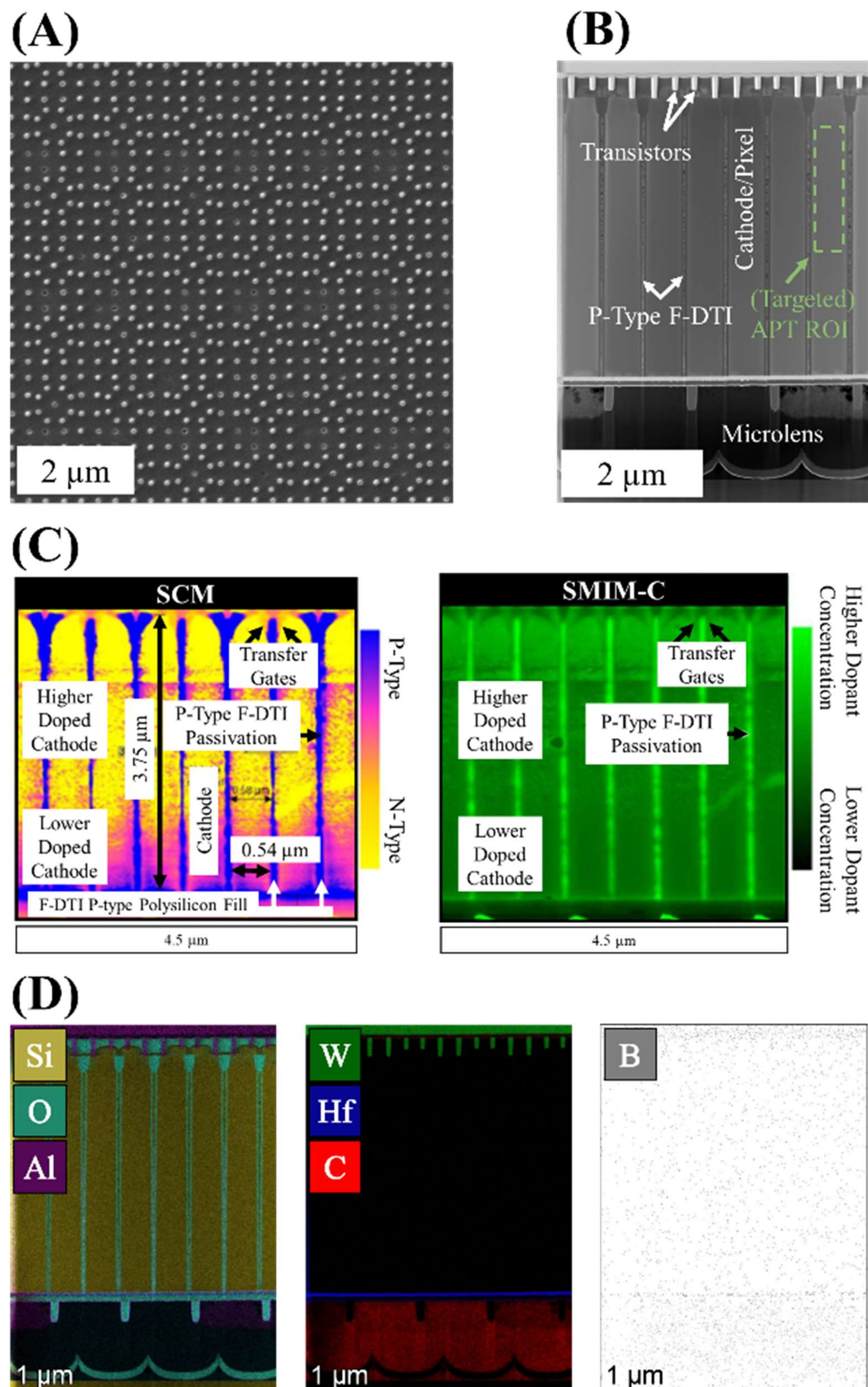

Figure 1. (A) Top-down SEM image of the CMOS image sensor at the polysilicon/contact level. (B) Cross-section HAADF-STEM image. (C) SCM and SMIM-C maps. (D) TEM-EDS elemental maps.

The FIB-SIMS instrument used in this study was the Raith IONMASTER magSIMS, integrating a tri-beam liquid metal alloy ion source (LMAIS) (i.e., Li, Ga, and Bi) with a magnetic sector mass spectrometer unit for surface analysis, including depth profiling, 2D and 3D imaging capabilities (Ost et al., 2024). Compared to analytical (S)TEM techniques, FIB-SIMS is particularly advantageous in semiconductor applications as it can differentiate isotopes and detect trace concentrations. Positive ion collection mode was used in order to maximize the extraction sensitivity of B and Si (Benninghoven et al., 1987); B was the primary dopant of interest. Of the three LMAIS ions, Bi was selected due to the low penetration depth (Nadzeyka et al., 2023), high sputtering yield, and reactivity (i.e., enhanced secondary ion generation) with dopants in semiconductors (Kim et al., 2019; Klump et al., 2018).

512×512 pixel maps (8 µm field-of-view) were acquired using a 35 kV $Bi^+$ beam (Figure 2(A); 5 µA emission, 13 pA primary current, 10 µm aperture, 2 ms/pixel). The acquired mass spectrum (Figure 2(A)) shows various identified peaks labelled accordingly. Other elements were also detected but not included (e.g., Na, Ti, Hf, W). Figure 2(B) shows the total secondary ion image of the device, while Figure 2(C) and Figure 2(D) provide the intensity maps of Si and B, respectively.

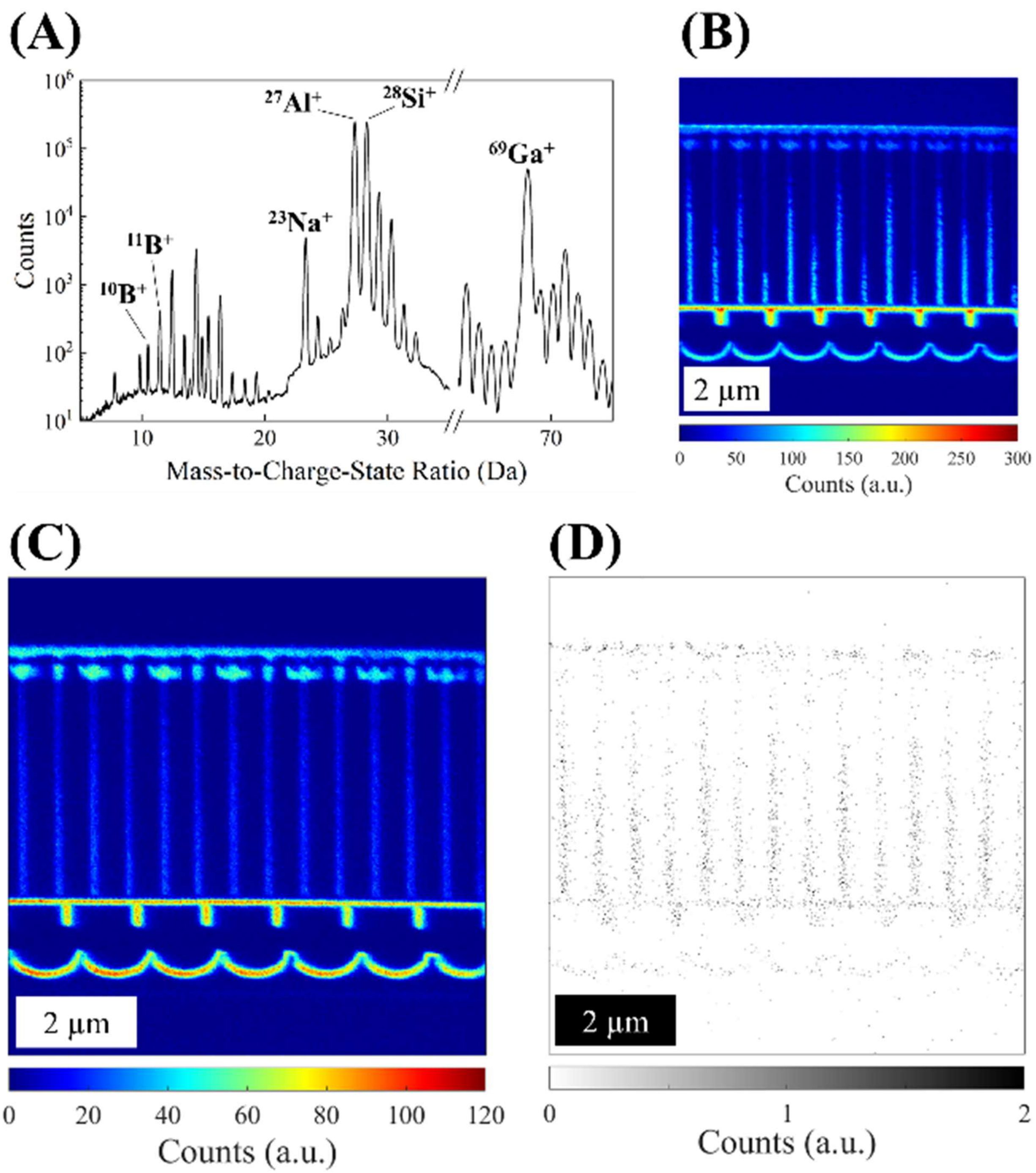

Figure 2. (A) FIB-SIMS mass spectrum. (B) Total secondary ion image. Intensity maps of (C) Si and (D) B.

## Results and Discussion

As indicated by the SCM map (Figure 1(C)), B distribution is generally isolated to the *p*-type doped polysilicon in the F-DTI. Within the pixels, however, there is negligible B detection.

Such a result may be due to the particularly low concentration(s) and the SMIM-C map (Figure 1(C)) does suggest that the *p*-type doped polysilicon in the F-DTI is more highly doped than the polysilicon in the pixels. Additionally, surface charge accumulation, which can distort ionization and collection efficiency (Wang et al., 2023), may be a contributing factor as reflected in the similarly low Si (and total) counts collected within the pixels.

A semi-quantitative evaluation of the depth profile is given by the integrated intensity plot in Figure 3, which shows the integrated B signal intensity as a function of pixel depth (summed over 3 image planes to enhance the signal-to-noise ratio). To validate signal integrity, a channel range (where no species are expected) was selected arbitrarily and included to estimate the background level. There is an initial sharp peak in the profile, indicative of highly-localized doped regions beneath the metallic contacts. Following this, the B profile is characterized by a broad distribution, increasing monotonically until a depth of ~5.5 μm. The two subsequent peaks are artifacts of an increased background level, which is due to the organic composition of the microlenses; organic structures have a propensity for fragmentation under ion bombardment (Gilmore, 2013; Popczun et al., 2017).

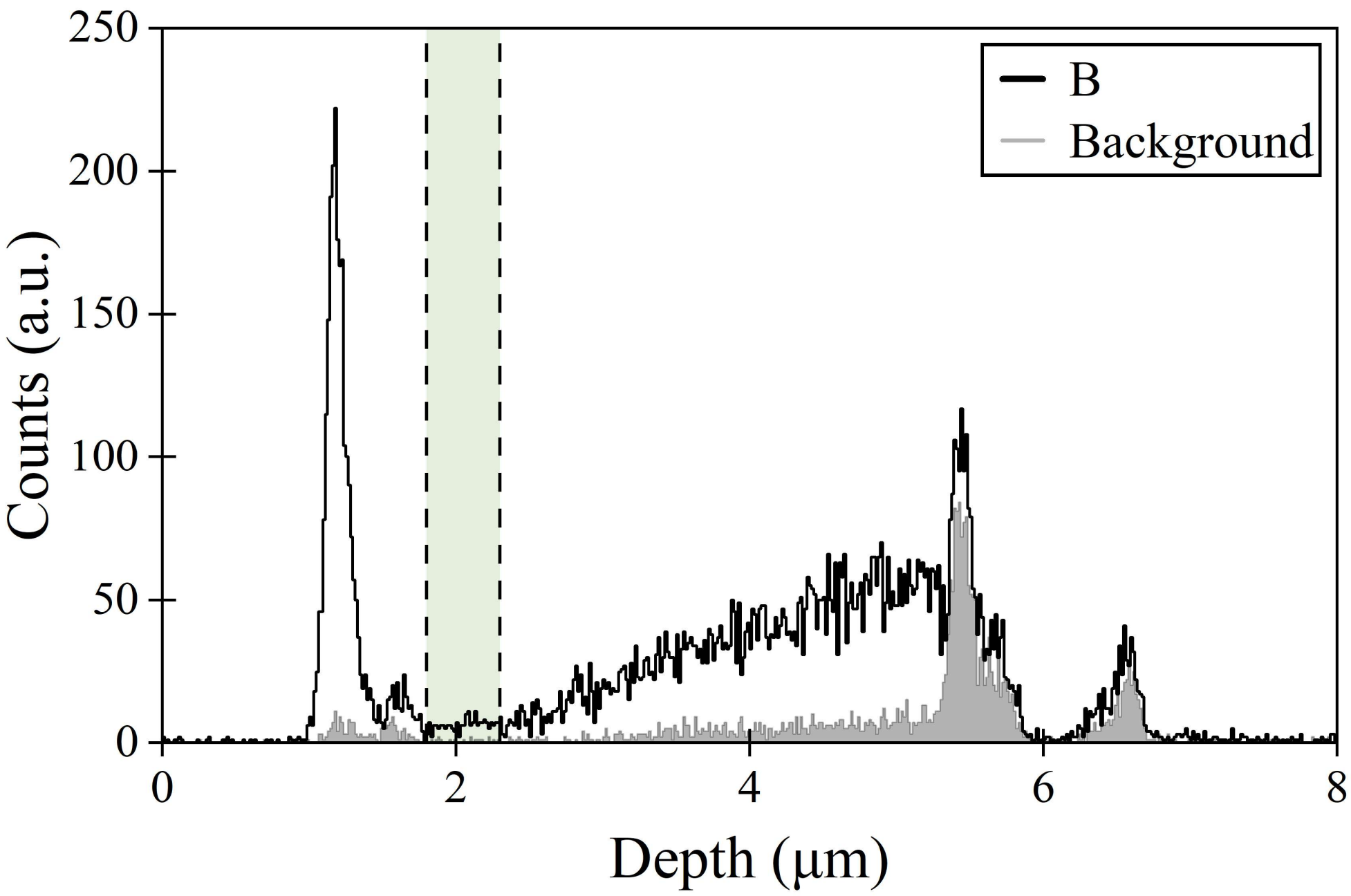

Figure 3. B integrated signal intensity plot (summed over 3 image planes; 512 pixel binning). The green shaded region indicates the analyzed depth of the sample using APT.

Without the use of a standard reference material for calibrating the signal intensities with atomic concentrations (Simons et al., 2007; Priebe et al., 2020), SIMS results are largely qualitative. Therefore, while FIB-SIMS can provide large-area dopant maps, APT is particularly advantageous for extracting quantitative, nanoscale information from a subset of the device. As indicated in Figure 1(B), APT analysis targeted regions within the pixels (not the F-DTI), as a very minute amount of B was detected there using FIB-SIMS. A starting depth of ~1.5-2 µm was chosen to reduce likelihood of premature fractures, on account of the W in the metallic contacts with a relatively high FE threshold. This is also the approximate starting position of the B profile in the *p*-type doped polysilicon within the F-DTI, as measured using FIB-SIMS (Figure 3). Specimens were prepared using the standard sample preparation protocol for APT needles (Thompson et al., 2007) on a Thermo Scientific Helios 5 UC DualBeam ($Ga^+$ source), ensuring a minimal shank angle in order to maximize the total depth of analysis. Samples exhibiting a large shank angle (e.g., > 10-15°) require a sustained voltage rise to maintain a constant detection rate, which may lead to the end-radius becoming too large to continue with analysis; LEAP instruments have a maximum voltage of ~12 kV (Larson et al., 2013).

APT was conducted in laser-mode on a LEAP 5000 XS, a straight flight-path instrument with 80% detection efficiency, at a base temperature of 50 K and 0.5% detection rate. For accurate measurements of B, a high electrostatic field (i.e., a low laser pulse energy (LPE) or a high Si charge-state ratio ($Si^{2+}/Si^{+}$)) has been shown to enhance measurement accuracy (Martin & Yatzor, 2019; Guerguis et al., 2024). Thus, a LPE of 16 pJ was used; further attempts to apply a lower value resulted in rapid specimen fractures, likely due to the high oxide content in the F-DTI. Data reconstruction was performed using the AP Suite software, with the radius evolution based on an SEM image of the specimen acquired after final sample preparation. When possible (i.e., no specimen fracture), the measured evaporated depth was used to calibrate the image compression factor using pre- and post-analysis specimen images.

Figure 4 shows an APT reconstruction and mass spectrum representative of the datasets collected. Oxide from a portion of the F-DTI was captured in the reconstruction but excluded from the analysis as it was not of interest. Notably, a minute As peak was detected in the mass spectrum and seen within the first 50 nm of the reconstruction. This peak, along with B and Si, is labelled accordingly in the mass spectrum, which also includes several molecular ions and $^{69}Ga^+$ from sample preparation. A bulk B concentration of ~$1.41\times10^{18}$ at/cm$^3$ was measured. This corresponds to ~30 parts per million and may further explain why B within the pixels was not detected using FIB-SIMS. Indeed, this is approaching the maximum sensitivity of the technique.

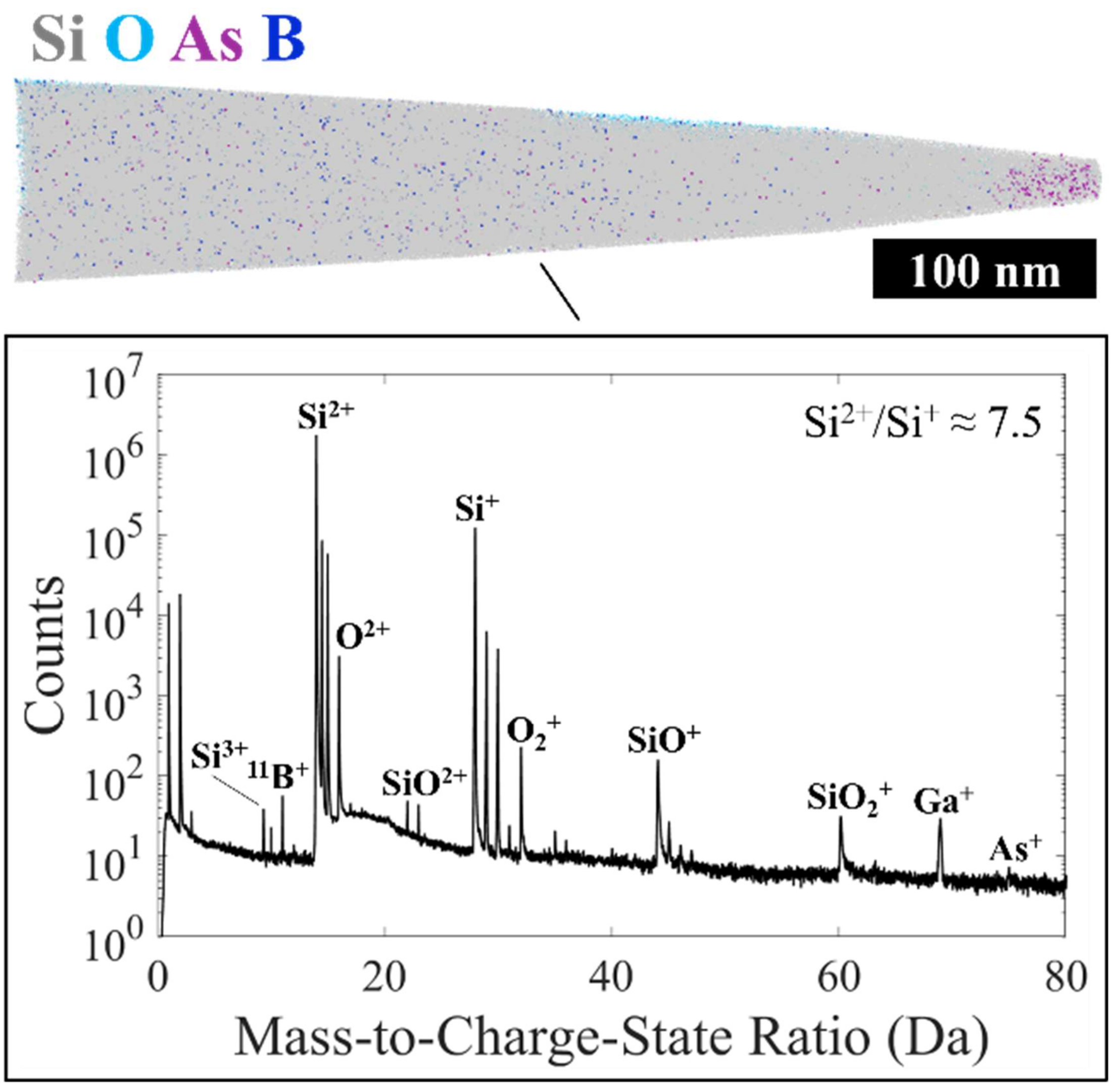

Figure 4. APT reconstruction and mass spectrum from a portion of the pixel. Oxide from the F-DTI was also captured in the reconstruction (opposite side) but excluded from analysis.

Bulk-volume As and B depth profiles were calculated using a 5 nm constant bin width (Figure 5). Regions on the mass spectrum adjacent to the B and As peaks were used to estimate the background level and correct the depth profiles accordingly. Based on a Si charge-state ratio of ~7.5, it is estimated that B was measured with ~80% accuracy (Guerguis et al., 2024); a 20% signal loss due to B's tendency for co-evaporation in multiple-hit burst events (Da Costa et al., 2012) and detector deadtime (Meisenkothen et al., 2015). As such, a correction was added to the profile to account for this factor (also in Figure 5). There is an initially high As concentration then rapidly dropping at ~50 nm, which is expected as the ROI begins beneath a vertical transfer gate. The vertical transfer gates (Figure 1(C)) are *n*-type doped polysilicon, and it is possible that the As detected is associated with (a portion of) this structure. B, however, increases monotonically over the 500 nm of analyzed depth. These results are in good agreement with previous SCM/SMIM-C mapping. In other words, an initially *n*-type doped (As) region followed by a gradient doped *p*-type (B) region with increasing concentration as a function of depth.

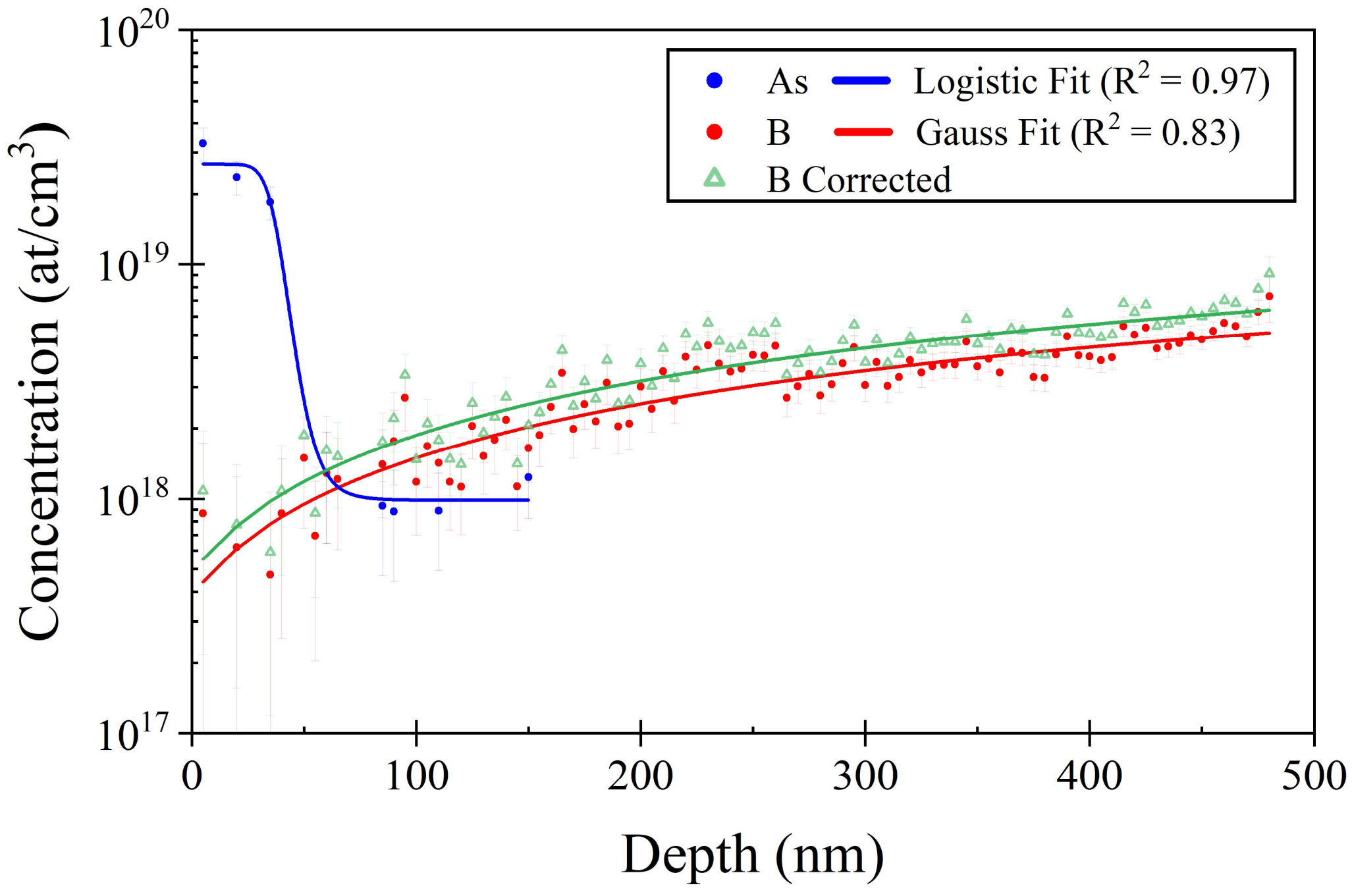

Figure 5. Background-corrected, bulk-volume As and B APT depth profiles with a 5 nm constant bin width.

## Conclusion

Thus, this letter has demonstrated the efficacy of applying four correlative techniques (i.e., SCM/SMIM-C, TEM, FIB-SIMS, and APT) for investigating semiconductor devices, bridging the gap between the micro- and nano- scales. Expanding this workflow to include additional FIB-SIMS imaging in negative ion collection mode (for assessing P and As distributions) would provide for a more holistic characterization of the device. Additionally, integrating this instrument with an '$O_2$-flooding' system (i.e., in-situ $O_2$ introduction to the sample surface during analysis) provides a future pathway to enhance secondary ion generation for a better dopants collection efficiency, as has been previously demonstrated (Franzreb et al., 2004; Magee et al., 2007).

To summarize, no *single* technique currently exists capable of addressing the challenges of quantifying trace dopant concentrations in advanced semiconductor devices:

- TEM has an unmatched spatial resolution, but a sensitivity that is insufficient for measuring low doping concentrations.
- SCM/SMIM-C can provide nanoscale electrical details with high sensitivity, but without any direct chemical information.
- FIB-SIMS can deliver large-area dopant maps with sub-20 nm resolution and a high throughput, but without known reference samples for quantification/calibration, information collected is largely qualitative.
- APT has an excellent combined spatial resolution and detection limit, but with stringent sample preparation requirements, faces challenges in analyzing certain material systems/architectures, and has a limited analysis volume (i.e., 50−100×50−100×100−500 $nm^3$).

Collectively however, these techniques are synergistic tools, enabling a comprehensive and holistic characterization of dopants in micro-devices.

## Acknowledgements

The authors thank the Natural Sciences and Engineering Research Council of Canada (NSERC) for supporting this work under the Discovery Grant program. APT and TEM work was carried out at the Canadian Centre for Electron Microscopy (CCEM), a national facility supported by McMaster University, the Ontario Research Fund (ORF), and the Canada Foundation for Innovation (CFI). B.G. is thankful to Gabe Arcuri and Travis Casagrande of the CCEM for assistance with sample preparation/APT analysis, along with TechInsights for providing the image sensor and conventional SIMS analysis.

## Data Availability

The data that support the findings of this study are available from the corresponding author upon reasonable request.

## Declarations of Competing Interests

One author (A.O.) is employed by Raith GmBH, which manufactures the FIB-SIMS IONMASTER instrument used in this work.

# PRE-PRINT